\documentclass[american,aps,pra,reprint, superscriptaddress]{revtex4-2}
\usepackage{color}
\usepackage{times}
\usepackage{subfig}
\usepackage{ dsfont }
\usepackage{bm,color,bbm}
\usepackage{graphicx}
\usepackage{graphics}
\DeclareGraphicsExtensions{.pdf,.png,.jpg}
\usepackage{amsbsy}
\usepackage{amsmath}
\usepackage{amsfonts}
\usepackage{amsthm}
\usepackage{float}
\usepackage[normalem]{ulem}
\usepackage{amssymb}
\usepackage{textcomp}
\usepackage[subnum]{cases}
\usepackage{multirow}
\usepackage{booktabs}
\graphicspath{{fig/}}

\definecolor{purple(html/css)}{rgb}{0.5, 0.0, 0.5}

\newcommand{\id}{\mathbbm{1}}

\newcommand{\stf}{\mathcal{F}}

\newenvironment{abbreviations}{\begin{list}{}{}}{\end{list}}

\begin{document}

\title{Trade-off between Bagging and Boosting for quantum separability-entanglement classification}

\author{Sanuja~D.~Mohanty}
\affiliation{Dept.~of~Physics, International~Institute~of~Information~and~Technology, Bhubaneswar 751029 Odisha, India}
\author{Ram~N.~Patro}
\email{c116009@iiit-bh.ac.in}
\affiliation{Dept.~of~ECE, International~Institute~of~Information~and~Technology, Bhubaneswar 751029 Odisha, India}
\author{Pradyut~K.~Biswal}
\affiliation{Dept.~of~ECE, International~Institute~of~Information~and~Technology, Bhubaneswar 751029 Odisha, India}
\author{Biswajit~Pradhan}
\affiliation{Dept.~of~Physics, International~Institute~of~Information~and~Technology, Bhubaneswar 751029 Odisha, India}
\author{Sk Sazim}
\email{sk.sazimsq49@gmail.com}
\affiliation{RCQI, Institute of Physics, Slovak Academy of Sciences, 845 11 Bratislava, Slovakia}
\affiliation{Center for Theoretical Physics, Polish Academy of Sciences, Aleja Lotnik\'{o}w 32/46, 02-668 Warsaw, Poland}

\begin{abstract}
Certifying whether an arbitrary quantum system is entangled or not, is, in general, an NP-hard problem. Though various necessary and sufficient conditions have already been explored in this regard for lower dimensional systems, 
it is hard to extend them to higher dimensions. Recently, an ensemble bagging and convex hull approximation (CHA) approach (together, BCHA) was proposed and it strongly suggests employing a machine learning technique for the separability-entanglement classification problem. However, BCHA does only incorporate the balanced 
dataset for classification tasks which results in lower average accuracy. In order to solve the data imbalance problem in the present literature, an exploration of the Boosting technique has been carried out, and a trade-off between the Boosting and Bagging-based ensemble classifier is explored for quantum separability problems. For the two-qubit and two-qutrit quantum systems, the pros and cons of the proposed random under-sampling boost CHA (RUSBCHA) for the quantum separability problem are compared with the state-of-the-art CHA and BCHA approaches. As the data is highly unbalanced, performance measures such as overall accuracy, average accuracy, F-measure, and G-mean are evaluated for a fair comparison. The outcomes suggest that RUSBCHA is an alternative to the BCHA approach. Also, for several cases, performance improvements are observed for RUSBCHA since the data is imbalanced.
\end{abstract}
\maketitle

\tableofcontents

\section{Introduction}
\label{sec:1}
Nowadays, machine learning (ML) is being employed more to tackle and solve harder problems in quantum information science. In recent years, it has been applied in state classifications \cite{lu2018separability,Harney_2020, PhysRevResearch.3.033278}, state reconstruction \cite{PhysRevLett.127.140502}, parameter estimation \cite{PhysRevA.100.062334}, and many others \cite{10.1038-s41534, Zhang-Xiao-Ming, Porotti-Ricca_commP, PhysRevX.8.031086, PhysRevA.103.L040401,rstoiuCristina, Schuff_2020,2022arXiv220109134L}. The motivation behind using ML in quantum information is to get more insights into problems where usual numerical techniques either fail or need more resources, eg., the optimization tasks in high constraint or non-convex scenarios.  

To decide whether an arbitrary quantum state is entangled or not is an NP-hard problem \cite{10.1145/780542.780545}. It is one of the long-standing fundamental issues in entanglement theory. A state of a composite system $\rho_{AB}$ is said to be separable if $\rho_{AB} = \sum_i p_i\rho_{A}^i \otimes \rho_{B}^i$ for any two subsystems $A$ and $B$, where $p_i$ $(\geq 0)$ represents classical mixing probability with $\sum_ip_i=1$. Otherwise, it is an entangled state. There exist numerous criteria to detect bipartite entanglement, however, these criteria are less reliable for higher dimensional systems. For example, the popular Peres-Horodecki criteria state that the separable states are positive partial transpose (PPT) \cite{peres1996separability,Horo-1996}, meaning for separable states $\rho_{AB}^{T_A}\geq 0$, where $T_A$ denotes transposition on system $A$. The criteria are necessary and sufficient for $d_Ad_B\leq 6$, where $d$ denotes system dimension. Other extant method includes entanglement witness, reduction criteria, cross-norm, or realignment criteria to name a few \cite{RevModPhys.81.865}. The most powerful technique is $k$-extension hierarchy, but it is notoriously hard to compute due to its exponentially growing complexity with $k$ \cite{PhysRevLett.88.187904, PhysRevA.80.052306}. Recently, in 
Ref.\cite{lu2018separability}, it was studied that ML techniques are instrumental in probing separability-entanglement classification. It was established that the ML-based technique is more efficient in terms of speed and accuracy than all extant methods. A couple more ML-based techniques were well studied for quantum separable-entanglement classification using artificial neural networks \cite{harney2021mixed, girardin2022building}.

Ref. \cite{lu2018separability} employed the convex hull approximation (CHA) to probe the separability-entanglement boundary using a supervised learning scheme. To reduce the error in classification using CHA, the bagging method \cite{10.1007/BF00058655} was invoked. This new method is known as bagging CHA (BCHA). This method increases the speed and accuracy of data manipulation as it divides the whole process into smaller units, and then runs in parallel. Ref. \cite{lu2018separability} demonstrates their results for two-qubits and two-qutrit systems with fairly high accuracy. 

In this work, building on the approaches of Ref. \cite{lu2018separability}, we propose an alternative method that addresses some important issues with further accuracy improvements for the separability-entanglement classification using ML. First, a) we notice that the earlier work doesn't address the issue of handling data imbalance, and b) did not explore all extant performance measures in their study. 

\section{Setting up the stage}
\label{sec:2}

\subsection{Supervised learning}
\label{sec:2a}

Supervised learning is a method of developing artificial intelligence that involves training a computer algorithm on input data that has been labeled for a certain output \cite{patro2019dictionary}. In order to apply it to real-time data, the model is trained until it can discover the underlying patterns and relationships between the input data and the output labels, allowing it to produce accurate classification results.

For supervised learning, the system is supplied with labeled data sets throughout its training phase, which tell it what output is associated with each specific input set. The trained model is then evaluated with test data, which is labeled data with the labels hidden from the algorithm \cite{patro2019conditional}. Further, the unlabeled testing data is used to determine how well the algorithm performs the classification task \cite{kotsiantis2007supervised}.

To create the learning dataset, we consider bipartite quantum state $\rho_{AB}$ of dimension $d_A\otimes d_B$ in $\mathcal{H}_A\otimes\mathcal{H}_B$. 
Arbitrary density matrix $\rho_{AB}$ $\in \mathcal{H}_A\otimes\mathcal{H}_B$ can be represented by real vector ${\bm x_i}$ $\in \mathcal{V}$ ($=\mathbb{R}^{d_A^2d_B^2-1}$) as $\rho^\dagger=\rho$ and ${\rm Tr}[\rho]=1$. We call such a vector \emph{feature vector} [see Appendix \ref{appen1} for detail]. 
The training dataset is then defined as $\Omega_{\rm train}=\{(\bm  x_i,y_i)|i=1,\cdots n\}$, where $x_i$ is the $i^{th}$ sample and $y_i$ is its corresponding class label, which is represented as, $y_i=1(0)$ if it is separable (entangled). Data labeling for $d_Ad_B\leq 6$ is performed by using PPT criteria. However, for higher dimensions, the labeling is done as per the Appendix-C of Ref. \cite{lu2018separability}.

In supervised learning, the main aim is to find a classifier (indicator function) $\Theta:  \mathcal{V}\to \{0,1\}$ which will fit the training data at best among a class of functions $\stf$.
As the present quantum entanglement is a binary classification problem, the error expresses the miss classification rate over two classes. For any training data $\Omega_{\rm train}$ consisting of $n$ samples, each associated with feature vector $\mathcal{V}$ and a target class label $y_i$ ($\in \{0,1\}$); the loss function $\mathbb{L}$ for any binary classifier $\Theta$ can be represented as
\begin{equation*}
\mathbb{L}(\Theta,\Omega_{\rm train})=\frac{1}{n}\sum_{i=1}^n \id [y_i \neq \Theta(\bm x_i)],
\end{equation*}
where $\id[\cdot]$ is a truth function of its argument. For any test data $\Omega_{\rm test}$, the value of function $\mathbb{L}(\Theta,\Omega_{\rm test})$ depicts the generalization error from $\Omega_{\rm train}$ to $\Omega_{\rm test}$.

It was found that among numerous extant supervised learning algorithms, eg., support vector machine (SVM) \cite{Cortes1995}, decision tree \cite{breiman1984}, boosting \cite{Schapire2003}, etc do not provide acceptable accuracy for separability problem \cite{lu2018separability}. This is due to the complex structure of the set of separable states. This led authors of Ref. \cite{lu2018separability} to the following consideration.

\subsection{Combining CHA with supervised learning}
\label{sec:2b}

The set of all separable states, $\Omega_1$, is convex and compact, and its exterior points are all pure product states. Using this fact, one can sample $\Omega_1$ using convex hull ($\mathds{C}$) of $m$ number of product states, $\{\bm c_i\}\in \mathcal{V}$, i.e., $\mathds{C}:={\rm conv}\{\bm c_i~|i=1,\dots, m\}$. The $\mathds{C}$ is the CHA of $\Omega_1$, and one can decide if an unknown state $\rho$ is separable or not by examining whether its feature vector $\bm x$ is in $\mathds{C}$. Equivalently, it is the solution of following linear programming:
\begin{align}
    &\max   ~ \alpha   ~~~ {\rm s.t}.  ~~~\alpha \bm x  \in \mathds{C},~~~{\rm i.e.},\nonumber\\
&\alpha \bm x = \sum_{i=1}^{|\mathds{C}|}\lambda_i\bm c_i, ~~ \lambda_i\geq 0, ~~\sum_{i}\lambda_i=1,
\label{LP1}
\end{align}
where $\alpha$ has functional dependence on both $\mathds{C}$ and $\bm x$. If $\bm x$ is in  $\mathds{C}$, then the corresponding state, $\rho$, is separable, else $\rho$ is an entangled state with high possibility. More specifically $\rho$ is separable when $\alpha\geq 1$ and entangled otherwise. We denote a maximal $\alpha$ for a chosen $m$-value as $\alpha_{\max}^m$. If we increase $m$ (to better approximate $\mathds{C}$), we will achieve better classification. It is evident that adding more exterior points in convex approximation will increase the accuracy of the above algorithms, however, it is really time-consuming. To overcome this, Ref. \cite{lu2018separability} used CHA in combination with supervised learning. Now, training data is defined as $\Omega_{\rm train}=\{(\bm x_i,\alpha_i,y_i)|i=1,\dots, n\}$ and the loss function of classifier $\Theta$ is redefined as 
\begin{equation}
\label{eq:L}
\mathbb{L}(\Theta,\Omega_{\rm train})=\frac{1}{n}\sum_{i=1}^n \id [y_i \neq \Theta(\bm x_i,\alpha_i)].
\end{equation}
Where $\alpha_i$ is the outcome of CHA for $i$-th random density matrix after solving the linear programming for finding $\bm x$ in $\mathds{C}$. Note that, CHA uses a threshold $\alpha \geq 1$ to classify as 1(0). The values of $\alpha$ acts as another feature for the classifier to learn the model. In Ref \cite{lu2018separability} bagging-based classification is performed on this feature space, known as Bagging CHA (BCHA). More information on the Bagging and Boosting approaches is discussed further.

\subsection{Overview of Bagging and Boosting Classifiers}
\label{sec:2c}

An ensemble meta-estimator called a bagging classifier fits base classifiers one at a time to random subsets of the original dataset, and then it aggregates the individual predictions (either by voting or by averaging) to provide a final prediction. By adding randomization to the process of building a black-box estimator (such as a decision tree), a meta-estimator of this kind can often be used to lower the variance of the estimator.

\begin{figure*}[!tb]
	\centering
	\includegraphics[scale=1]{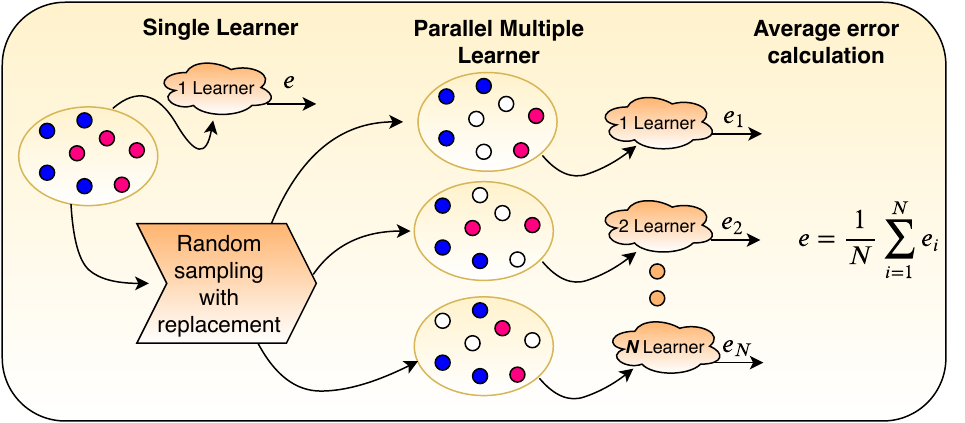}
	\caption{Overview of Bagging classifier: Multiple learners are created by generating additional data points. The new data points are created randomly with a uniform probability as before. Generally, the created $N$ learners are parallel and are further averaged to obtain the final learning error defined as $e=\frac{1}{N}\sum_{i=1}^{N} e_i$.}
\label{fig:bag}
\end{figure*}

A training set is created by randomly selecting $M$ instances (or pieces of data) from the original training dataset (of size $N$), 
and used to train each base classifier in parallel. Each base classifier's training set is distinct from the others. In the resultant training set, many of the original data might be replicated while others might not. An overview of Bagging classifiers is presented in Fig. \ref{fig:bag}.

\begin{figure*}[!tb]
	\centering
	\includegraphics[scale=1]{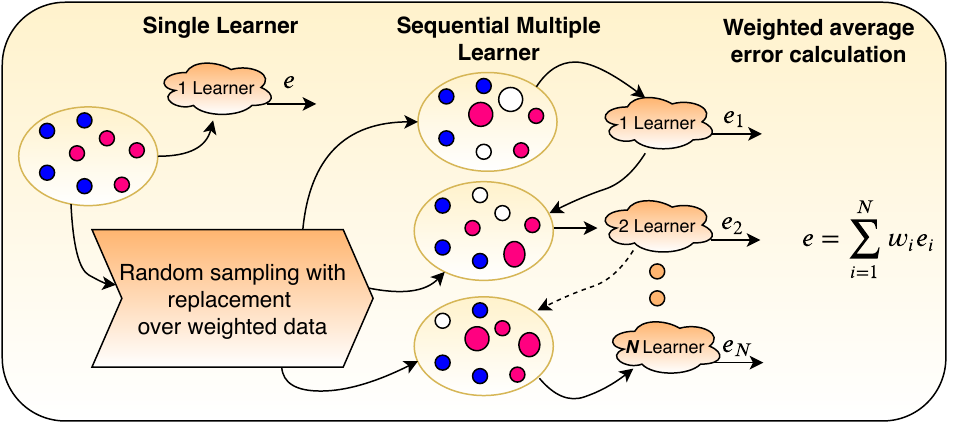}
	\caption{Overview of Boosting classifier: Similar to the Bagging approach, the Boosting classifier also generates multiple data points. But, unlike parallel in Bagging, the Boosting approach sequentially learns the error from the previous learner and assigns a higher weight to the miss classified data, and random sampling with weighted  replacement is carried out. Also, another set of weights assigned to the learners} are further accumulated to find the final weighted average error defined as $e=\sum_{i=1}^{N} w_i e_i$.
	\label{fig:boost}
\end{figure*}

A number of weak classifiers are combined in the broad ensemble approach known as "boosting" to produce a strong classifier. In order to do this, a model is first constructed using the training data, and a second model is then developed in an effort to fix the errors in the first model. The training set is predicted exactly or a predetermined number of models are added, depending on which comes first. AdaBoost \cite{schapire2013explaining} was the first really successful boosting algorithm developed for binary classification. An overview of Boosting classifiers is presented in Fig. \ref{fig:boost}.

Both boosting and bagging fall under the category of "ensemble learning." Combining many weak learners to create a hybrid categorization system. Most often, "ensemble learning" refers to trained weak decision ensemble trees.

\subsection{Imbalanced dataset}
\label{sec:2d}

\textit{Imbalanced dataset} refers to an unequal distribution of class samples within a dataset. Such unequal distribution of class samples reduces the training performance of the classifiers, and hence the classification results on the testing data are also affected. 

In the present context, the volume of entangled states is far more than the separable states, making the dataset imbalanced. For more details on the experimented datasets, see Section \ref{sec:4a}. From the discussion in Section \ref{sec:4a}, we can observe that the prevalence differences are high for both datasets and hence they are highly imbalanced.

This demands a classifier that can handle data imbalance issues and can be more suitable for quantum separability-entanglement classification problems. Which is discussed in the next section. 

Also, for such imbalanced datasets, the learning performance of any ML approach is greatly affected \cite{japkowicz2002class} and needs a careful performance evaluation. Such performance measures are discussed in Section \ref{sec:4b}.

\subsection{Ensemble classifiers for imbalanced dataset}
\label{sec:2e}

It has been well studied that, for imbalanced data, the SVM classifier may be biased towards the majority class \cite{akbani2004applying}. A modification of SVM has already been presented, incorporating random under-sampling (RUS) for an unbalanced dataset \cite{tang2008svms} by removing the samples randomly from the training set. For highly unbalanced data, synthetic minority oversampling technique (SMOTE) \cite{chawla2002smote, 10.1007/978-981-13-1592-3_48} has been applied towards classification, where, it generally over-sample the minority class to create synthetic data points. So further incorporation of SMOTE to Boosting approach may be effective for classification. When oversampling is performed by duplicating examples,
it may lead to over-fitting \cite{drummond2003c4}. So, further modification by incorporating the under-sampling may help in the performance improvement of the classifier. Instead of over-sampling the minority classes, under-sampling the majority classes also may help in improving the classifier results. The RUS randomly removes examples from the majority class until the desired class
distribution is found \cite{seiffert2009rusboost}. Such integration with Boosting is RUSBoost \cite{seiffert2009rusboost}, which is a hybrid approach combining random under-sampling, SMOTE, and Adaptive Boost (AdaBoost) classifier.

For ensemble learning, bagging and boosting are generally applied (see Fig. \ref{fig:bag} and Fig. \ref{fig:boost}). Already the Bagging-based CHA (BCHA) is proposed \cite{lu2018separability}, reporting higher accuracy than CHA. But, as the data is highly unbalanced, the accuracy evaluation should be twofold -- 1) Overall accuracy (OA) and 2) Average accuracy (AA). For more details on the performance measures OA and AA, see \ref{sec:4b}. OA is the number of correctly classified test samples per total samples under test. While AA is the sum of accuracy for each class predicted per the total number of classes (average of each accuracy per class). Hence, although the reported OA \cite{lu2018separability} is higher, we evaluated the AA of BCHA, which is of less margin than the CHA approach. This demands further improvement in the classifier which can take care of both the OA and AA for separability-entanglement classification. 

As the experimented data set is highly unbalanced (refer Section \ref{sec:4a}), the RUSBoost approach is explored for separability-entanglement classification and is validated over the state-of-the-art approaches. The subsequent section describes the RUSBoost ensembled CHA classifier.

\section{RUSBoost CHA (RUSBCHA)}
\label{sec:3}

Initially, all examples in the training data set are assigned equal weights. During each iteration of AdaBoost, a weak hypothesis is formed by the base learner. The error associated with the hypothesis is calculated, and the weight of each example is adjusted such that wrongly classified examples have their weights increased while correctly classified samples have their weights decreased. Therefore, subsequent iterations of boosting will generate hypotheses that are more likely to correctly classify the previously mislabeled examples. After all, iterations are completed, a weighted vote of all hypotheses is used to assign a class to the unlabeled samples.

Data sampling techniques attempt to alleviate the problem of class imbalance by adjusting the class distribution of the training data set. This can be accomplished by either removing examples from the majority class (under-sampling) or adding examples to the minority class (oversampling).

SMOTE adds new artificial minority examples by extrapolating between preexisting minority instances rather than simply duplicating original examples. The newly created instances cause the minority regions of the feature space to be fuller and more general.

The RUSBoost takes advantage of all these approaches by combining them. A detailed discussion on the RUSBoost approach can be found in \cite{seiffert2009rusboost}.

Although significant classifier performance improvement is observed \cite{lu2018separability} in the case of BCHA as compared to standalone CHA, some limitations exist which are discussed in Section \ref{sec:1}. So, it can be further improvised in two ways 1) by replacing the classifier and 2) by increasing the feature space by proper feature extraction technique. Presently the first case is explored by incorporating the RUSBCHA classifier for possible improvement in the classification results leaving scope to explore the feature extraction techniques as future work.

\section{Experimental Setup}
\label{sec:4}
All the classifications were carried out on two kinds of feature spaces 1) vector represented $\rho$ ($d^2-1$ dimensional feature space), 2) vector represented $\rho$ with CHA calculated $\alpha_{\max}^m$ for a specific $m$ ($d^2$ dimensional feature space).
The experiments are carried out for both the two-qubit and two-qutrit systems. Five different techniques such as; Bagging, Boosting were tested on raw $d^2$-1 (for two-qubit system $d$=4 and for two-qutrit system $d$=9) dimensional feature vector  $\bm{x}$, CHA with only one $\alpha_{\max}^m$,  while, the BCHA and RUSBCHA are trained with both the $\bm{x}$, and $\alpha_{\max}^m$. Their associated feature spaces are presented in Table \ref{tab:feat}. 
\begingroup

\begin{table}[htbp]
	\centering
		\begin{tabular}{@{}cccccc@{}}
			\toprule
			& BAGGING       & BOOSTING      & CHA & BCHA        & RUSBCHA     \\ \hline
			Feature space & $d^2$-1 & $d^2$-1 & $d^2$-1            & $d^2$ & $d^2$ \\
			Two qubit     & 15                     & 15                     & 15            & 16                   & 16                   \\
			Two qutrit   & 80                     & 80                     & 80            & 81                   & 81                   \\ \hline
		\end{tabular}
		\caption{Various experimented classifiers with their associated feature space (dimensions).}
	\label{tab:feat}
\end{table}
\endgroup
The dataset details and the performance evaluators are presented below.

\subsection{Dataset preparation}
\label{sec:4a}

\begin{figure}[!htb]
	\centering
	\includegraphics[scale=0.3]{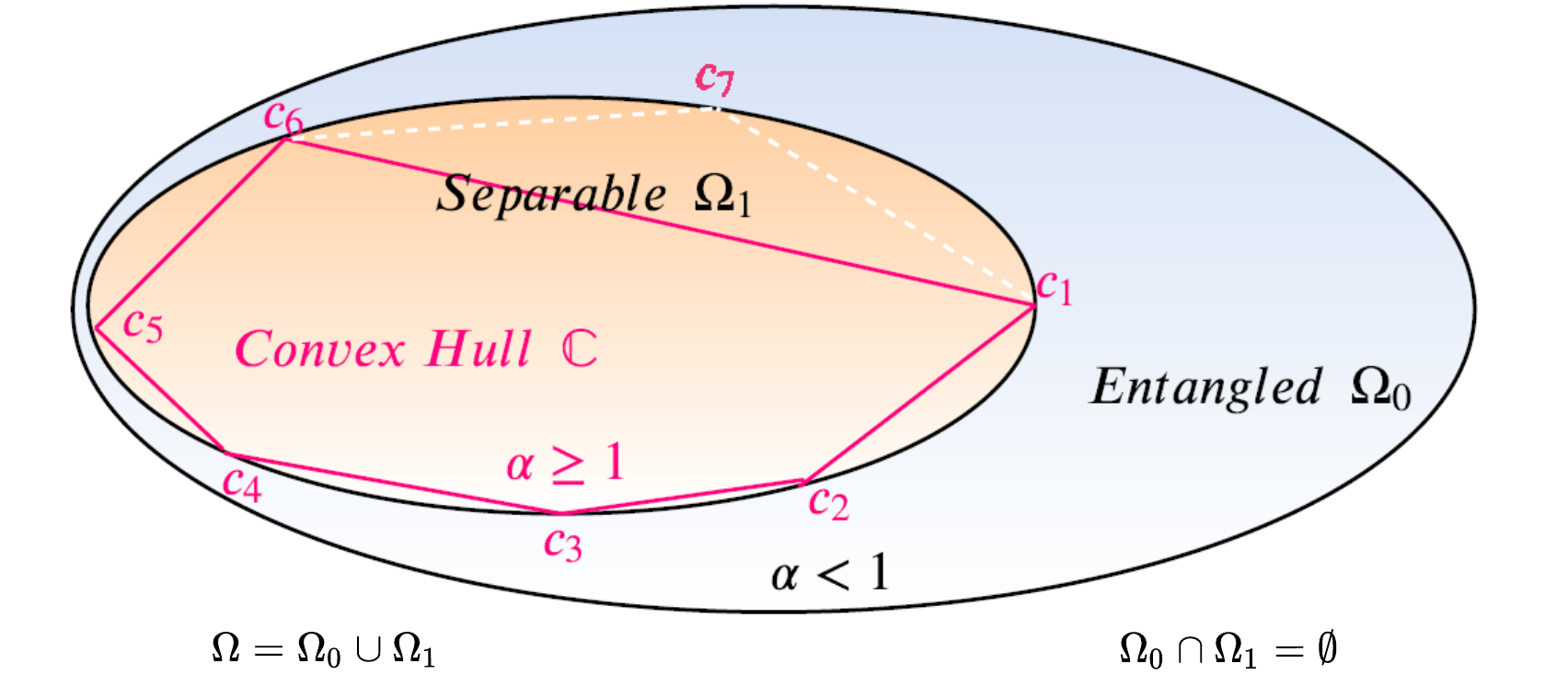}
	\caption{Data space $\Omega$ as a combination of entangled $\Omega_{0}$ and separable $\Omega_{1}$ subspaces. $c_i$ represents the pure product states.}
	\label{fig:space}
\end{figure}

The total data space $\Omega$ is a combination of the separable subspace $\Omega_{1}$ and entangled subspace $\Omega_{0}$; such that $\Omega=\Omega_{1}\cup \Omega_{0}$ and $\Omega_{1}\cap \Omega_{0}=\emptyset$ (see Fig \ref{fig:space}). Two datasets, 
representing the feature vectors of random density matrices for two-qubit and two-qutrit systems respectively, are supplied with their class labels in \cite{qmlab}. The procedure for creating the random separable and entangled states can be referred to in the BCHA manuscript \cite{lu2018separability}. The total and class-specific training and testing sample information for the pair of the experimented datasets; namely two-qubit and two-qutrit system, are presented in Table \ref{tab:data22} and Table \ref{tab:data33} respectively. Approximate 50\% samples are randomly selected for training and the remaining 50\% samples are used for testing to evaluate the performances of ML algorithms.

The maximized parameter, $\alpha_{\max}^m$ for CHA (with varying $m$) of 1) two-qubit system with $m=[1000,2000,..,10000]$, and 2) two-qutrit system with $m=[10000,20000,..,100000]$) were also obtained from \cite{lu2018separability,qmlab}. The minimization was made by solving the linear programming defined in Eq.(\ref{LP1}).

\begin{table}[htbp]
	\centering
\begin{tabular}{@{}c|c|c|c@{}}
	\toprule
	Class (label) & Total & Training & Testing \\ \hline
	Separable (1) & 2814           & 1388              & 1426             \\
	Entangled (0) & 37186          & 18612             & 18574            \\
	All sample    & 40000          & 20000             & 20000            \\ \hline
\end{tabular}
\caption{Dataset description of experimented training, testing, and total samples for two-qubit systems.}
	\label{tab:data22}
\end{table}

\begin{table}[htbp]
	\centering
		\begin{tabular}{@{}c|c|c|c@{}}
			\toprule
			Class (label) & Total & Training & Testing \\ \hline
			Separable (1) & 6751           & 3338              & 3413             \\
			Entangled (0) & 13249          & 6662              & 5687             \\
			All sample    & 20000          & 10000             & 10000            \\ \hline
		\end{tabular}
		\caption{Dataset description of experimented training, testing, and total samples for two-qutrit systems.}
	\label{tab:data33}
\end{table}

From Table \ref{tab:data22} and Table \ref{tab:data33}, we can observe that the class samples are unequally distributed within the dataset. A prevalence difference for a binary classification represents the degree of imbalance in the dataset. The dataset-specific prevalence difference of class samples can be interpreted as, for:

\begin{itemize}
    \item Two-qubit dataset (Table \ref{tab:data22}) : $\left| \frac{2814}{40000} - \frac{37186}{40000}\right|=0.8593$.
    \item Two-qutrit dataset (Table \ref{tab:data33}): $\left| \frac{6751}{20000} - \frac{13249}{20000}\right|=0.3249$.
\end{itemize}

For a balanced dataset, the prevalence difference must approach $0$. However, we can observe that the prevalence difference for the two-qubit dataset is high (0.86) and for the two-qutrit dataset, it is comparatively low (0.32). This clearly signifies that the experimented dataset is highly imbalanced. For such imbalanced datasets, the learning performance of any ML approach is greatly affected \cite{japkowicz2002class} and needs a careful performance evaluation. Such performance measures are discussed further.

\subsection{Performance measures}
\label{sec:4b}

\begin{figure}[!htb]
	\centering
	\includegraphics[scale=0.2]{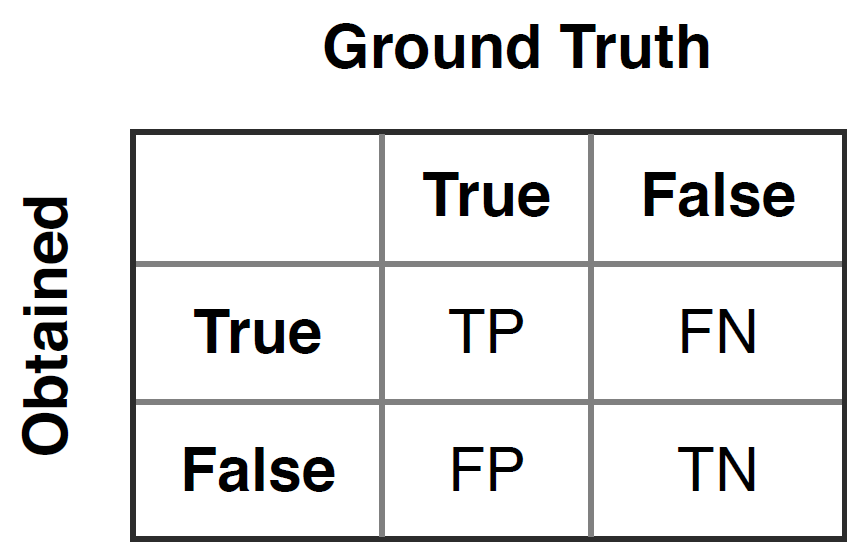}
	\caption{Confusion matrix for binary classification.}
	\label{fig:ConM}
\end{figure}

For ease of understanding the binary classification, the confusion matrix is presented in Fig. \ref{fig:ConM}. In the figure, columns represent the original class labels (supplied with the data) as \emph{true} and \emph{false}, similarly each row represents the outcome of the classifier. 

True positive (TP) and true negative (TN) are defined as both the original (ground truth) and the obtained (classified) class labels are true and false respectively. The contradictions are presented as false positive (FP) and false negative (FN) which are off-diagonal in the confusion matrix. Let $N$ number of samples be tested, i.e., $N=\sum \left( TP+TN+FP+FN\right)$. So, higher TP and TN values lead to better accuracy; on the contrary, higher FP and FN values reject the classifier.

Now we can define overall accuracy (OA) as
\begin{equation*}
 OA = \frac{TP+TN}{N},
\end{equation*}
and the overall error (OE) as $OE=1-OA$.

For binary classification, let, out of $N$ tested samples, there are $N_1$ and $N_2$ samples labeled as \emph{true} and \emph{false} respectively, (where $N=N_1+N_2$). The average accuracy (AA) is the mean accuracy obtained for each class and is defined as
\begin{equation*}
 AA= \frac{1}{2} \left( \frac{TP}{N_1}+\frac{TN}{N_2} \right).
\end{equation*}
and the average error (AE) as $AE=1-AA$.

Similarly, other important measures such as sensitivity ($s=\frac{TP}{TP+FN}$),
specificity ($r=\frac{TN}{N}$),
Precision ($k= \frac{TP}{TP+FP}$),
F-measure  and
G-mean can be incorporated for validating the classification results. We will use the following two for our analysis: 
\begin{equation*}
\mbox{F-measure} = 2\left(\frac{k \times s}{k + s}\right),\:\:\mbox{and}\:\:
%
\mbox{ G-mean} = \sqrt{s \times r}.
\end{equation*}
Higher values of OA, AA, F-measure, and G-mean are desirable for evaluating the performance of a classifier.

\section{Results and Discussion}
\label{sec:5}

We used both the datasets (see Section \ref{sec:4a}) and all the performance measures described in Section \ref{sec:4b}, to compare the proposed RUSBCHA and other state-of-art classifiers in terms of figures. For the robust representation of performances on the experimented data, all the classification performance measures are averaged over 30 independent evaluations.

\begin{figure}[!htb]
	\centering
	\includegraphics[scale=0.6]{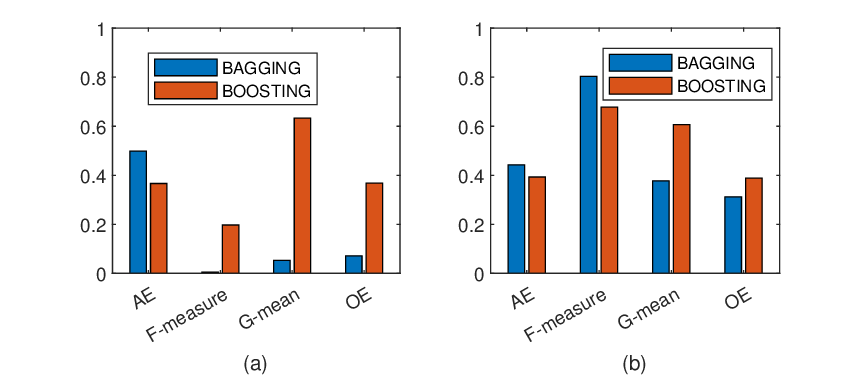}
	\caption{Classification results of the raw data without considering the CHA ($\alpha$) for (a) two-qubit and (b) two-qutrit system.}
	\label{fig:no_alpha}
\end{figure}

The Bagging and Boosting classifier only incorporates the $d^2$-1 dimensional feature vector $\bm{x}$. The classification performance as; AE, F-measure, G-mean, and OE; for two-qubit and two-qutrit systems are presented in Fig. \ref{fig:no_alpha} (a) and Fig. \ref{fig:no_alpha} (b) respectively. For the two-qubit system, (Fig. \ref{fig:no_alpha} (a)) it is observed that the proposed Boosting approach outperforms the Bagging approach in terms of F-measure, G-mean, and AE. While marginal deviation is observed for OE. Similarly, for the two-qutrit system (Fig. \ref{fig:no_alpha} (b)), improvement is observed for G-mean and AE. 
According to both the CHA and BCHA approaches, if $\alpha_{\max}^m \geq 1$, $\bm{x}$ is separable; else, $\bm{x}$ is highly possible to be an entangled state. Hence, our proposed RUSBCHA classifier also incorporates both the feature vectors $\bm{x}$ and $\alpha_{\max}^m$. To find the trade-off between the state-of-the-art BCHA and the proposed RUSBCHA approach, further experiments are made on both two-qubit and two-qutrit datasets. These experiments include: 

\begin{itemize}
    \item \textbf{Experiment 1:} Performance evaluation of classifiers over varying $m$.
    \item \textbf{Experiment 2:} Performance evaluation of classifiers over varying percentages of training and testing samples.
    \item \textbf{Experiment 3:} Performance evaluation of classifiers on varying prevalence difference of dataset.
\end{itemize}

\subsection{Experiment 1}
\begin{figure}[!htb]
	\centering
	\includegraphics[width = 0.5 \textwidth]{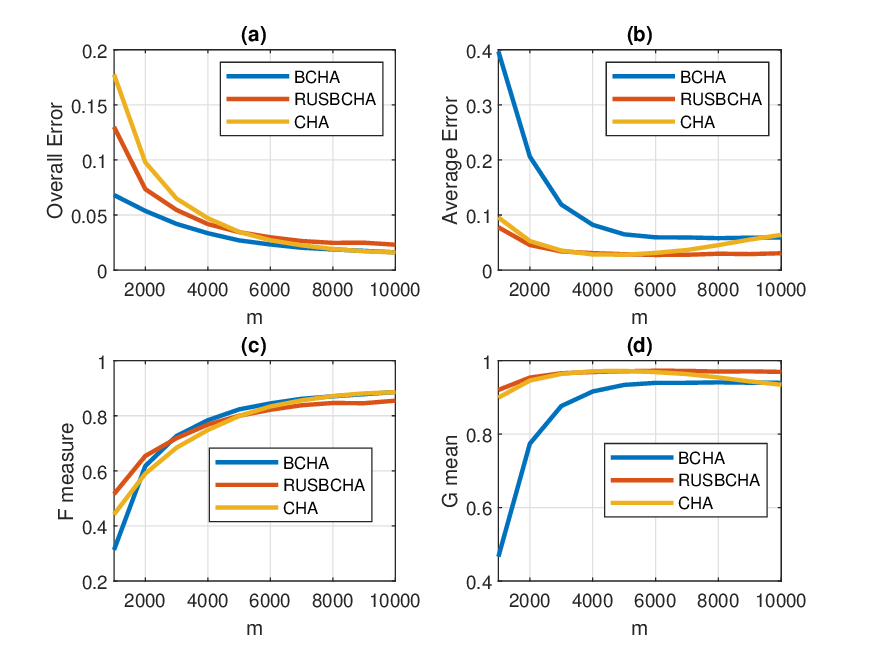}
	\caption{Classification results of the two-qubit system, considering the CHA ($\alpha$).}
	\label{fig:alpha_22rdm}
\end{figure}

\begin{figure}[!htb]
	\centering
	\includegraphics[width = 0.5 \textwidth]{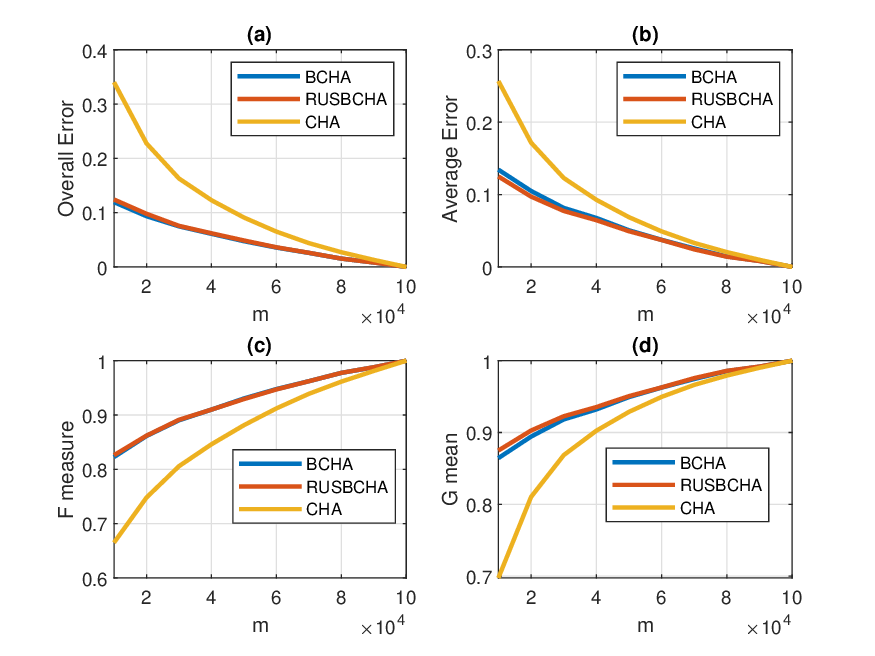}
	\caption{Classification results of the two-qutrit system, considering the CHA ($\alpha$).}
	\label{fig:alpha_33rdm}
\end{figure}

In this experiment, the CHA, BCHA, and proposed RUSBCHA classifiers are compared over varying $m$ for both two-qubit and two-qutrit datasets. Experimental results are shown in Fig. \ref{fig:alpha_22rdm} and Fig. \ref{fig:alpha_33rdm}.

For a two-qubit system, from the Fig. \ref{fig:alpha_22rdm}(b), it can be observed that the AE of BCHA is higher for all values of $m$ as compared to CHA and RUSBCHA approaches. The BCHA performance has almost 40\% error for the lower value of $m$. It can also be observed that, for lower values of $m$, the performances of CHA and RUSBCHA are similar, while, for higher values of $m$ RUSBCHA has lower AE values. This clearly signifies that the proposed RUSBCHA is less biased to the majority classes and hence the average accuracy is higher in comparison to other state-of-approaches. A similar interpretation also can be seen in Fig. \ref{fig:alpha_22rdm}(d).

From the Fig. \ref{fig:alpha_22rdm}(a), it can be observed that the OE of BCHA has lower values, and hence its performance is better for lower values of $m$ in comparison to RUSBCHA and CHA approaches. While the proposed RUSBCHA has intermediate performance in comparison to other state-of-approaches. However, in Fig. \ref{fig:alpha_22rdm}(c), the F-measure performances are equivalently similar for all approaches.

On the other hand, for the two-qutrit system (Fig. \ref{fig:alpha_33rdm}), both the BCHA and RUSBCHA have similar performances over varying $m$ with significant performance improvements as compared to the state-of-art CHA approach.

In this experiment, you can observe better performance of proposed RUSBCHA approach for two-qubit dataset in comparison to BCHA and CHA approaches. While similar performances are observed for both RUSBCHA and BCHA for two-qutrit datasets. To find the rationale for performance differences of these two datasets, further experiments are carried out.

\subsection{Experiment 2}
\begin{figure}[!htb]
	\centering
	\includegraphics[scale=0.63]{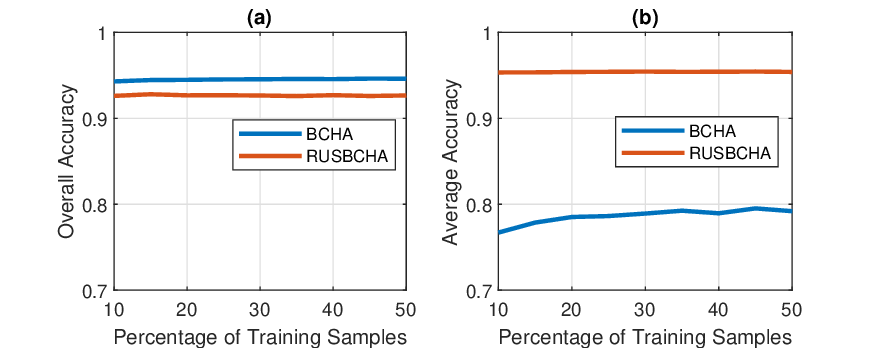}
	\caption{Obtained Overall  Accuracy and Average Accuracy for the two-qubit system over varying percentage (\%) of training samples ($m$=2000).}
	\label{fig:compare1}
\end{figure}

\begin{figure}[!htb]
	\centering
	\includegraphics[scale=0.63]{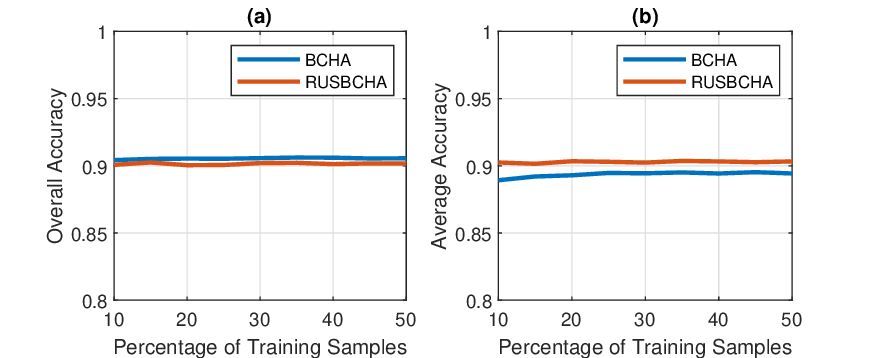}
	\caption{Obtained Overall  Accuracy and Average Accuracy for the two-qutrit systems over varying percentage (\%) of training samples ($m$=20000).}
	\label{fig:compare2}
\end{figure}

In literature, it is proved that several machine learning techniques such as neural network and deep learning require a large number of samples to train. The above problem may occur due to the sensitivity of the classifier to the percentage of training samples. In experiment 1, $50$\% of samples are trained and the rest are tested. Hence, further validation of the approaches is carried out with varying training ($10$\%-$50$\%) and testing (90\%-50\%) scales, and the performances are presented in Fig. \ref{fig:compare1} and Fig. \ref{fig:compare2} for two-qubit and two-qutrit systems respectively. Note that, for this experiment, the total samples are the same as Table \ref{tab:data22} and Table \ref{tab:data33} for the respective datasets. In this experiment, $m$ is set as 2000 and 20000 for two-qubit data and two-qutrit data respectively.

 From the Fig. \ref{fig:compare1}(a), it can be observed that OA of BCHA is 2.5\% more than RUSBCHA, while in the Fig. \ref{fig:compare1}(b) AA of RUSBCHA is more than 15\% better than BCHA. However, the results of these classifiers do not vary by the variation in training percentages. Therefore, performance of both the classifiers is not sensitive to the number of training samples. For the two-qutrit data, in Fig. \ref{fig:compare2}(a) and Fig. \ref{fig:compare2}(b), you can also observe similar results. However, the AA performances in Fig. \ref{fig:compare1}(b) and Fig. \ref{fig:compare2}(b) suggests that the RUSBCHA performs better than BCHA, specifically for two-qubit dataset. Note in this respect that the prevalence difference of the two-qutrit dataset ($0.3249$) which is comparatively low referring to the prevalence difference of the two-qubit dataset ($0.8593$) for this experiment. This further suggests that doing further experiments to test both the classifiers with varying prevalence difference ratios might provide us some clue on how these classifiers work for imbalanced datasets.

\subsection{Experiment 3}
The above experiments were performed with two-qubit and two-qutrit datasets as mentioned in Table \ref{tab:data22} and Table \ref{tab:data33} respectively. From these tables, you can observe that the separable samples are only 7\% and 33\% of the total samples for two-qubit and two-qutrit datasets, respectively. To test the performance of classifiers for different prevalence differences, we created imbalanced datasets of different prevalence differences for both two-qubit and two-qutrit. 

\begin{table}[h]
\caption{Description of imbalanced datasets created from the original two-qubit dataset of Table \ref{tab:data22}.}
\label{tab:data222}
\begin{tabular}{@{}cccc@{}}
\toprule
\multicolumn{3}{c}{\textbf{Number of Samples}}           & \multicolumn{1}{l}{\multirow{2}{*}{\textbf{\begin{tabular}[c]{@{}l@{}}Prevalence \\ Difference\end{tabular}}}} \\ \cmidrule(r){1-3}
\textbf{Separable} & \textbf{Entangled} & \textbf{Total} & \multicolumn{1}{l}{}                                                                                           \\ \midrule
1800               & 37186              & 38986          & 0.907                                                                                                          \\
2814               & 33000              & 35814          & 0.842                                                                                                          \\
2814               & 18000              & 20814          & 0.729                                                                                                          \\
2814               & 14000              & 16814          & 0.665                                                                                                          \\
2814               & 10000              & 12814          & 0.560                                                                                                          \\
2814               & 8000               & 10814          & 0.479                                                                                                          \\
2814               & 5500               & 8314           & 0.323                                                                                                          \\
2814               & 4500               & 7314           & 0.230                                                                                                          \\
2814               & 3500               & 6314           & 0.108                                                                                                          \\
2814               & 3000               & 5814           & 0.031                                                                                                          \\ \bottomrule
\end{tabular}
\end{table}
Table \ref{tab:data222} shows the description of created imbalanced datasets for two-qubits. In this table, each row describes a dataset which is a subset of the dataset described in Table \ref{tab:data22}. For each created dataset subset, its number of separable, entangled, and total samples are represented. Also for each entry in the table, the prevalence difference of the respective dataset is mentioned. One notices the prevalence difference values range approximately from $0$ to $0.9$. The value $0$ represents the dataset is balanced, and value $0.9$ represents the dataset is highly imbalanced. A similar interpretation for the two-qutrit dataset can be done from the Table \ref{tab:data333}.

\begin{figure}[!htb]
	\centering
	\includegraphics[scale=0.5]{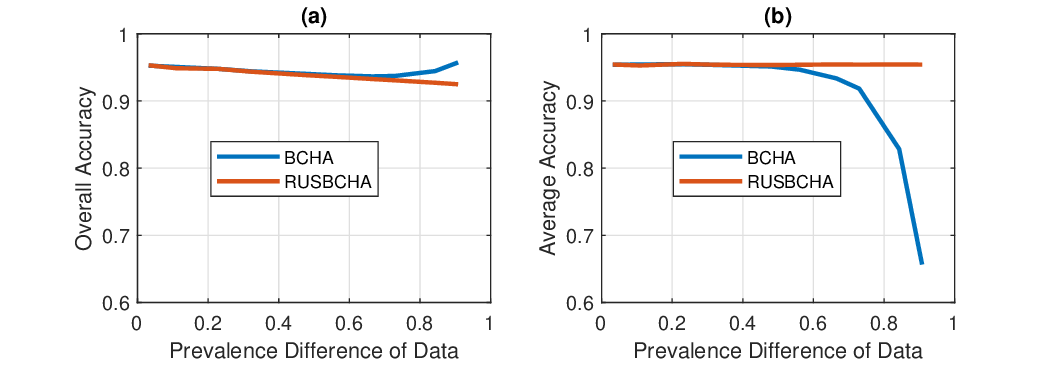}
	\caption{Overall Accuracy and Average Accuracy of BCHA and RUSBCHA over varying prevalence difference of the two-qubit data.}
	\label{fig:preValence2}
\end{figure}

\begin{table}[h]
\caption{Description of imbalanced datasets created from the original two-qutrit dataset of Table \ref{tab:data33}.}
\label{tab:data333}
\begin{tabular}{cccc}
\hline
\multicolumn{3}{c}{\textbf{Number of Samples}}           & \multicolumn{1}{l}{\multirow{2}{*}{\textbf{\begin{tabular}[c]{@{}l@{}}Prevalence \\ Difference\end{tabular}}}} \\ \cline{1-3}
\textbf{Separable} & \textbf{Entangled} & \textbf{Total} & \multicolumn{1}{l}{}                                                                                           \\ \hline
600                & 13249              & 13849          & 0.913                                                                                                          \\
1380               & 13249              & 14629          & 0.811                                                                                                          \\
2200               & 13249              & 15449          & 0.715                                                                                                          \\
3200               & 13249              & 16449          & 0.610                                                                                                          \\
4200               & 13249              & 17449          & 0.518                                                                                                          \\
5500               & 13249              & 18749          & 0.413                                                                                                          \\
6751               & 13000              & 19751          & 0.316                                                                                                          \\
6751               & 10500              & 17251          & 0.217                                                                                                          \\
6751               & 8500               & 15251          & 0.114                                                                                                          \\
6751               & 7000               & 13751          & 0.018                                                                                                          \\ \hline
\end{tabular}
\end{table}
Fig. \ref{fig:preValence2} shows the classifier performances over the varying prevalence of two-qubit data. In the figure, the performances are averaged over 30 iterations, and in each iteration, a new subset of the dataset is created with varying prevalence differences (Table \ref{tab:data222}). For this experiment, we fixed these parameters $m$=2000, and 50\% training samples.

It is observed from the Fig. \ref{fig:preValence2}(a) that the OA of both BCHA and RUSBCHA are similar up to 0.6 prevalence difference. However, afterward, there is a minor improvement of OA for BCHA approach in comparison to RUSBCHA approach. From the Fig. \ref{fig:preValence2}(b) it can be observed that both BCHA and RUSBCHA performances are similar up to 0.5 prevalence difference. However, afterward, there is a sharp decline of AA for BCHA in comparison to RUSBCHA.

\begin{figure}[!htb]
	\centering
	\includegraphics[scale=0.5]{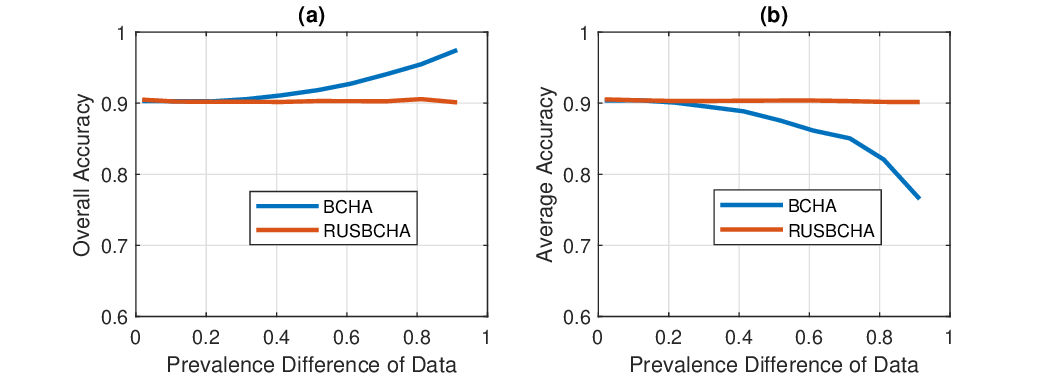}
	\caption{Overall Accuracy and Average Accuracy of BCHA and RUSBCHA over varying prevalence difference of the two-qutrit data.}
	\label{fig:preValence3}
\end{figure}

Fig. \ref{fig:preValence3} shows the classifier performances over the varying prevalence of two-qutrit data. In the figure, the performances are averaged over 30 iterations, and in each iteration, a new subset of the dataset is created with varying prevalence differences (Table \ref{tab:data333}). For this experiment, we fixed these parameters $m$=20000, and 50\% training samples.

It is observed from the Fig. \ref{fig:preValence3}(a) that the OA of both BCHA and RUSBCHA are similar up to 0.3 prevalence difference. However, afterward, there is a minor improvement of OA for BCHA approach in comparison to RUSBCHA approach. From the Fig. \ref{fig:preValence3}(b) it can be observed that both BCHA and RUSBCHA performances are similar up to 0.25 prevalence difference. However, afterward, there is a sharp decline of AA for BCHA in comparison to RUSBCHA.
From the results in Fig. \ref{fig:preValence2} and Fig. \ref{fig:preValence3}, it can be observed that the performance of the proposed RUSBCHA approach is consistent (almost a straight line) over varying prevalence differences of data. So, it can be concluded that the performance of RUSBCHA is not heavily affected by the data imbalances.

Referring to our earlier observations, for Fig. \ref{fig:alpha_22rdm}: the reason for having good AA of proposed RUSBCHA over BCHA; and for Fig. \ref{fig:alpha_33rdm}: the reason for having similar performances of both RUSBCHA and BCHA can now be justified using Fig. \ref{fig:preValence2} and Fig. \ref{fig:preValence3} respectively. Since the prevalence difference of two-qubit data is 0.8593 our proposed RUSBCHA performs better than BCHA. While the prevalence difference of two-qutrit data is 0.3249, hence, both BCHA and RUSBCHA performances are similar.

Hence, we can conclude that the RUSBCHA can be an alternative to the BCHA approach and also can be a better classifier to deal with highly imbalanced datasets. Overall, the ensemble learning is helpful for better understanding of separability-entanglement problem, when compared to the stand-alone CHA approach.

\section{Conclusion}
\label{sec:6}

The necessity of a separability-entanglement classifier is well-known in the quantum information forum. Although various necessary and sufficient criteria like PPT have been proposed in the past, still, they cannot be generalized for higher dimensions. The ML approaches are vastly exploited in the general data-mining perspective, while the discussions and applications are limited in quantum information processing. Similar to BCHA, we proposed RUSBCHA as an alternative ML-based solution for the quantum separability problem. 
The proposed RUSBCHA approach for quantum separability problem shown improvements in AE for the two-qubit system; while having similar responses for the two-qutrit systems in comparison to CHA. As the data is highly unbalanced, standard performance measures like OE, AE, F-measure, and G-mean are evaluated.  The results suggest incorporating a proper ML approach to classify the separability-entanglement criteria with proper performance matrices. Also, the proposed RUSBCHA can be an alternative to CHA which can deal with the unbalanced dataset that may reduce the over-fitting error of the classifier.

In order to evaluate the effectiveness of the classifier, the feature extraction is unexploited here, however, this can be a further direction of research to improve the classification performance. Also, other ML approaches can be exploited and validated further.

\begin{acknowledgements}
SS acknowledges funding through Pasific program call 2 (Agreement No. PAN.BFB.S.BDN.460.022 with the Polish Academy of Sciences). This project has received funding from the European Union’s Horizon 2020 research and innovation programme under the Marie Sk{\l}odowska- Curie grant agreement No 847639 and from the Ministry of Education and Science. SS also acknowledges the financial support through DEQHOST (APVV-22-0570) and DESCOM (VEGA-2/0183/21) during his stay at IPSAS, Bratislava. 
\end{acknowledgements}

\section*{Author declarations}
\noindent The authors have no conflicts to disclose.

\section*{Code and Data Availability Statement}
\noindent We have uploaded the code and the data created for our analysis in the following open GitHub repository: https://github.com/ram-patro/RUSBCHA. 

\noindent Note that the earlier repository, QMLab \cite{qmlab} created by the authors of Ref. \cite{lu2018separability} is no longer available. Our repository given above includes all of the analysis by QMLab also.

\appendix
\section{Feature vector}\label{appen1}
To illustrate what is feature vector $\bm x$, we consider the following example. We know a quantum state ($\rho_d$) in $d$-dimensional Hilbert space can be represented by a $d\times d$ density matrix using generalised Gell-Mann matrices,  $\sigma_i \in$ $SU(d)$ as 
\begin{align}
    \rho_d=\frac{1}{n}\left(\mathbb{I}+\sqrt{\frac{d(d-1)}{2}}\bm x. \bm \sigma\right),
\end{align}
where $\bm x\in \mathbb{R}^{d^2-1}$ is the feature vector which satisfies $x_i=\sqrt{\frac{d}{2(d-1)}}{\rm Tr}[\rho_d\sigma_i]$. This is possible as $\rho$ is Hermitian and has trace unity.

In our analysis, we consider quantum systems in $d_A\otimes d_B$ dimensional Hilbert space $\mathcal{H}_A\otimes\mathcal{H}_B$ which are represented by $d_A d_B\times d_A d_B$ density matrices. Hence, to represent using feature vectors, we need Gell-Mann matrices $\sigma_i\in SU(d_A d_B)$, i.e., the $\bm x\in \mathbb{R}^{d_A^2d_B^2-1}$.

\section{Generating random density matrices in the code}\label{appen2}
Most of the contents in the appendix are elaborately discussed in Ref.\cite{lu2018separability}. We will discuss the methods of producing random density matrices for specific dimensions in a nutshell. 

To produce random bipartite density matrices of any rank numerically, we use the probability distribution $p(\mu,\theta,d)=\mu\times\triangle_\theta$, where $\mu$ is the uniform distribution on $U(d)$ according to the Haar measure, $\triangle_\theta$ is the Dirichlet distribution 
\begin{align}
    \triangle_\theta(\ell_1,\cdots,\ell_d):=C_\theta\prod_{i=1}^d \ell_i^{-\theta}
\end{align}
defined on the simplex $\sum_i^d\ell_i=1$, where $\theta>0$ is a parameter and $C_\theta$ is a normalization constant. We set $\theta=\frac{1}{2}$ for sampling both the two-qubit and two qutrit states.

Note that our dataset is exactly the same as is used in Ref.\cite{lu2018separability}. The Ref.\cite{lu2018separability} observed the following trends during training using the generated samples: 
\begin{itemize}
    \item For the two-qubit case, approximately 7\% of the states among $5\times 10^4$ are PPT, i.e., separable state.
    \item Among fairly large samples (randomly generated) of two-qutrits, only 2.2\% are PPT. After rejecting all the states with negative partial transpose while sampling as they are assumed entangled (prior information), the total collected PPT states are a total of $2\times 10^4$ samples. Among PPT states, at least 66.24\% are found to be separable using CHA. However, note that during the testing, NPT states are also included.
\end{itemize}
The authors in Ref.\cite{lu2018separability} observe that these trends are consistent with the previously predicted ones in Ref.\cite{PhysRevA.60.3496}.
\vspace{2mm}

\section{List of Abbreviations}
\begin{abbreviations}
\item[CHA] Convex Hull Approximation
\item[BCHA] Bagging based CHA
\item[RUSBCHA] Random Under Sampling BCHA
\item[ML] Machine Learning
\item[PPT] Positive Partial Transpose
\item[SVM] Support Vector Machine
\item[SMOTE] Synthetic Minority Oversampling Technique
\item[TP] True Positive
\item[TN] True Negative
\item[FP] False Positive
\item[FN] False Negative
\item[OA] Overall Accuracy
\item[AA] Average Accuracy
\item[OE] Overall Error
\item[AE] Average Error

\end{abbreviations}

\end{document}